\documentclass[prc,twocolumn,epsfig,nofootinbib,floatfix]{revtex4}
\usepackage{graphics}
\usepackage{epsfig}
\usepackage{amsfonts}
\usepackage{amsmath}
\usepackage{bm}

\begin{document}
\title{
SKYRME-RPA DESCRIPTION OF SPIN-FLIP M1 GIANT RESONANCE}
\author{P. Vesely$^1$, J. Kvasil$^1$, V.O. Nesterenko$^2$, W. Kleinig$^{2,3}$,
P.-G. Reinhard$^4$, and V.Yu. Ponomarev $^5$}
\affiliation{$^1$ Institute of
Particle and Nuclear Physics, Charles University, CZ-18000, Praha 8, Czech
Republic}
\affiliation{$^2$ Laboratory of Theoretical Physics, Joint Institute
for Nuclear Research, Dubna, Moscow region, 141980, Russia}
\affiliation{$^3$ Technische Universit\"at
Dresden, Inst. f\"ur Analysis,D-01062, Dresden, Germany}
\affiliation{$^4$
Institut f\"ur Theoretische Physik II, Universit\"at Erlangen, D-91058,
Erlangen, Germany}
\affiliation{$^5$ Institut f\"ur Kernphysik, Technische
Universit\"at Darmstadt, D-64289 Darmstadt, Germany}
\date{\today}

\begin{abstract}
The spin-flip M1 giant resonance is explored in the framework of Random Phase
Approximation  on the basis of the Skyrme energy functional. A
representative set of eight Skyrme parameterizations (SkT6, SkM*, SLy6, SG2,
SkO, SkO', SkI4, and SV-bas) is used. Light and heavy, spherical and deformed
nuclei ($^{48}$Ca, $^{158}$Gd, $^{208}$Pb, and $^{238}$U) are considered. The
calculations show that spin densities play a crucial role in forming the
collective shift in the spectrum. The interplay of the collective shift and
spin-orbit splitting determines the quality of the description. None of the
considered Skyrme parameterizations is able to describe simultaneously the M1
strength distribution in closed-shell and open-shell nuclei. It is found that
the problem lies in the relative positions of proton and neutron spin-orbit
splitting. Necessity to involve the tensor and isovector spin-orbit interaction
is called for.
\end{abstract}


\maketitle

\section{Introduction}

Nuclear density-functional-theory (DFT), with the most prominent
representatives being Skyrme-Hartree-Fock (SHF), the Gogny forces, and the
relativistic mean-field model, achieved a high level of quality in the
description of ground state and dynamics of atomic nuclei
\cite{Ben,Vre05aR,Sto07aR}. Most applications for nuclear dynamics up to now
have been concerned with electrical excitation modes (natural parity). The
description works generally well, except for some persistent problems with the
isovector giant resonance (GR) in light nuclei \cite{Rei99a,svbas}.  Much less work
has been done yet for magnetic excitations (unnatural parity). At the same
time, magnetic modes are sensitive to a different class of force parameters,
namely those related to spin.  An exploration of magnetic resonances, like
spin-flip M1, could essentially improve the spin-orbit interaction. Magnetic
modes could clarify the role of still vague tensor interaction
\cite{Les_PRC_07_tensor}. Also, the spin-flip M1 resonance is
a counterpart of the spin-isospin Gamow-Teller resonance which is of great
current interest in connection with astrophysical problems
\cite{Vre05aR,Sto07aR,Bender_PRC_02_GT,Fracasso_PRC_07_GT,Sarrin_NPA_01_GT}.
Investigation of the M1 resonance could be useful in this connection as well.

There are many studies of the spin-flip M1 mode within simple models
beyond the DFT, see e.g. reviews \cite{Speth_91,Osterfeld_92,Harakeh_book_01}.
At the same time, as far as we know, a DFT treatment is
limited to a few publications using SHF
\cite{Hilton_98,Sarriguren_M1}  and even that is not carried through fully
consistently.  The work \cite{Sarriguren_M1} uses a hybrid model with partial
inclusion of SHF in the Landau-Migdal formulation \cite{Sarriguren_M1}. The
other study \cite{Hilton_98} uses the early Skyrme forces
and omits  the crucial spin density. These studies,
being a useful first step, are not satisfactory for nowadays demands.

The present study aims at a fully self-consistent description of the spin-flip
M1 mode in the framework of SHF. Previous investigations for Gamow-Teller
\cite{Bender_PRC_02_GT,Fracasso_PRC_07_GT,Sarrin_NPA_01_GT}
and spin-flip M1 modes \cite{Sarriguren_M1} hint that spin-density
response could be decisive to get a sizeable collective shift.
So all the Skyrme terms with spin density (usually omitted in
calculations for electric modes) have to be implemented and
scrutinized. Furthermore, because of the obvious importance of
spin-orbit splitting, the responses delivered by spin-orbit and tensor
interactions have to be inspected as well.
Since the quality of the description may depend on the particular
Skyrme parameterization as well as on nuclear shape and mass region, a
variety of parameterizations should be checked for light and heavy,
spherical and deformed nuclei.  Note that the M1 mode in heavy
open-shell nuclei (rare-earth and actinides) exhibits a pronounced
double-peak structure \cite{exp_Gd_U,exp_Gd} while
closed-shell nuclei ($^{48}$Ca, $^{208}$Pb, ...) show only one peak
\cite{exp_48Ca,exp_208Pb,exp_Shizuma_08}. All these demands are met in the present
study which, to the best of our knowledge, is the first systematic
and self-consistent SHF exploration of spin-flip M1.

\section{Details of calculations}

We will consider the resonance in the doubly-magic nuclei
$^{48}$Ca and $^{208}$Pb and  axially deformed nuclei $^{158}$Gd
and $^{238}$U.  A representative set of eight SHF parameterizations is
used: SkT6 \cite{skt6}, SkO \cite{sko}, SkO' \cite{sko}, SG2
\cite{sg2}, SkM* \cite{skms}, SLy6 \cite{sly46}, SkI4 \cite{ski3}, and
SV-bas \cite{svbas}. They exhibit a variety of effective
masses (from $m^*/m$=1 in SkT6 down to 0.65 in SkI4) and other nuclear
matter characteristics. Some of the forces (SLy6) were found favorable in
the description of E1(T=1) GR
\cite{nest_PRC_06,nest_IJMPE_07,nest_IJMPE_08,nest_PRC_08}.  Others
were used in studies of Gamow-Teller strength (SG2, SkO')
\cite{Bender_PRC_02_GT,Fracasso_PRC_07_GT,Sarrin_NPA_01_GT,sg2} or
peculiarities of spin-orbit splitting (SkI4) \cite{ski3}.  The
forces SkT6, SG2 and SkO' involve the tensor spin-orbit term.  The
parameterization SV-bas represents one of the latest upgrades in SHF.

The calculations are performed within the self-consistent
separable random-phase-approximation (SRPA) approach based on
factorized Skyrme residual interaction
\cite{nest_PRC_02,nest_05_lanl,nest_PRC_06}.  The self-consistent
factorization considerably reduces the computational expense of RPA
while maintaining a high accuracy. This allows to
perform systematic studies in both spherical and deformed (heavy
and super-heavy) nuclei
\cite{nest_PRC_02,nest_PRC_06,nest_IJMPE_07,nest_IJMPE_08,nest_PRC_08}.
The residual interaction includes all
contributions arising from the SHF functional as well as
the Coulomb (direct and exchange) and pairing (at BCS level) terms
\cite{nest_05_lanl,nest_PRC_06}).

The Skyrme energy density to be exploited reads \cite{Ben,Sto07aR}
\begin{eqnarray}
  \mathcal{H}_\mathrm{Sk}
  &=&
  \frac{b_0}{2} \rho^2- \frac{b'_0}{2} \sum_q\rho_{q}^2
  + \frac{b_3}{3} \rho^{\alpha+2}
  - \frac{b'_3}{3} \rho^{\alpha} \sum_q \rho^2_q
\nonumber
\\
 &&
 +b_1 (\rho \tau - \textbf{j}^2)
 - b'_1 \sum_q(\rho_q \tau_q - \textbf{j}^2_q)
\nonumber
\\
 &&
 - \frac{b_2}{2} \rho\Delta \rho
 + \frac{b'_2}{2} \sum_q \rho_q \Delta \rho_q
\nonumber
\\
 &&
 - b_4 (\rho \nabla\textbf{J}\!+\!(\nabla\!\times\!\textbf{j})\!\!\cdot\!\!\textbf{s})
\nonumber
\\
 &&
 - b'_4 \sum_q (\rho_q \nabla\textbf{J}_q\!
 +\!(\nabla\!\times\!\textbf{j}_q)\!\!\cdot\!\!\textbf{s}_q)
\nonumber
\\
  &&
  + \frac{\tilde{b}_0}{2} \textbf{s}^2
  - \frac{\tilde{b}'_0}{2} \sum_q \textbf{s}_{q}^2
+ \frac{\tilde{b}_3}{3} \rho^{\alpha} \textbf{s}^2
- \frac{\tilde{b}'_3}{3} \rho^{\alpha} \sum_q \textbf{s}^2_q
\nonumber
\\
  &&
 -\frac{\tilde{b}_2}{2} \textbf{s} \!\cdot\!
  \Delta \textbf{s} + \frac{\tilde{b}'_2}{2}
  \sum_q \textbf{s}_q \!\cdot\!\Delta \textbf{s}_q
\nonumber
\\
 &&
  +\gamma_\mathrm{T}\tilde{b}_1
   (\textbf{s}\!\cdot\!\textbf{T}\!-\!\textbf{J}^2)
  + \gamma_\mathrm{T}\tilde{b}'_1
   \sum_q (\textbf{s}_q\!\cdot\!\textbf{T}_q
    \!-\!\textbf{J}_q^2)
\label{eq:skyrme_funct}
\end{eqnarray}
where $b_i$,
$b'_i$, $\tilde{b}_i$, $\tilde{b}'_i$ are the force parameters.
This functional involves time-even (nucleon $\rho_q$, kinetic-energy
$\tau_q$, spin-orbit $\textbf{J}_q$) and time-odd (current
$\textbf{j}_{ q}$, spin $\textbf{s}_q$, vector kinetic-energy
$\textbf{T}_q$) densities where $q$ denotes protons and
neutrons. Densities without index, like $\rho = \rho_p + \rho_n$,
denote total densities. The contributions with $b_i$ (i=0,1,2,3,4) and
$b'_i$ (i=0,1,2,3) are the standard terms responsible for ground state
properties and electric excitations of even-even nuclei
\cite{Ben,Sto07aR}. In the standard SHF, the isovector
spin-orbit interaction is linked to the isoscalar one by
$b'_4=b_4$. The tensor spin-orbit terms
$\propto\tilde{b}_1,\tilde{b}'_1$ are often skipped. In (\ref{eq:skyrme_funct})
they can be switched by the parameter $\gamma_\mathrm{T}$.
The spin terms  with $\tilde{b}_i, \tilde{b}'_i$ become relevant only for odd
nuclei and magnetic modes in even-even nuclei.
Though $\tilde{b}_i, \tilde{b}'_i$ may be uniquely
determined as functions of ${b}_i,{b}'_i$ \cite{Sto07aR}, their values
were not yet well tested by nuclear data. Moreover,
following a strict DFT, they can be considered as free parameters. Just these
spin terms may be of a paramount importance for the spin-slip M1. Hence all them
are taken into account in SRPA.

In addition to second functional derivatives entering
the SRPA residual interaction for electric modes,
\begin{equation}\label{electr_contr}
\frac{\delta^2 E}{\delta \rho_{q'} \delta \rho_{q}} \; ,
\frac{\delta^2 E}{\delta \tau_{q'} \delta \rho_{q}} \; ,
\frac{\delta^2 E}{\delta \textbf{J}_{q'} \delta \rho_{q}}\; ,
\frac{\delta^2 E}{\delta \textbf{j}_{q'} \delta \textbf{j}_{q}} \; ,
\end{equation}
the present treatment also involves  the terms with
\begin{equation}\label{spin_contr}
\frac{\delta^2 E}{\delta \textbf{j}_{q'} \delta \textbf{s}_{q}} \; ,
\frac{\delta^2 E}{\delta \textbf{s}_{q'} \delta \textbf{s}_{q}} \; ,
\frac{\delta^2 E}{\delta \textbf{J}_{q'} \delta \textbf{J}_{q}} \; ,
\frac{\delta^2 E}{\delta \textbf{T}_{q'} \delta \textbf{s}_{q}} \; .
\end{equation}
SRPA generators include spin and orbital input operators
$\hat{P}^s_{q}=R(r)\hat{s}_+^q$ and  $\hat{P}^l_{q}=R(r)\hat{l}_+^q$ with
$R(r)$ being 1 or $r^2$. In deformed nuclei, to take into account the coupling
between spin and quadrupole $K^{\pi}=1^+$ states, the quadrupole generator
$\hat{Q}_{q}=r^2 Y_{21}(\Theta)$ is also added. The convergence of the results
with including more generators was checked. See details in \cite{Petr_PhD}.

The SHF calculations employ a coordinate-space grid with the mesh size
0.7 fm.  For deformed nuclei, cylindrical coordinates are used and the
equilibrium quadrupole deformation is found by minimization of the
total energy \cite{nest_PRC_06,nest_PRC_08}.
The single-particle spectrum involves all levels from
the bottom of the mean field well up to +20 MeV. In the heaviest
nucleus under consideration, $^{238}U$, this results in $\sim$17000
two-quasiparticle (2qp) $K^{\pi}=1^+$ pairs with the excitation
energies up to 50-70 MeV.  Note that for electric E1(T=1) and E2(T=0)
excitations such single-particle space provides a satisfying exhaustion of
the energy-weighted sum rules \cite{nest_IJMPE_08}.

The spectral distribution of the spin-flip M1 mode with $K^{\pi}=1^+$
is presented as the strength function
\begin{equation}
  S(M1 ; \omega) = \sum_{\nu \ne 0}
  |\langle\Psi_\nu|\hat{M}|\Psi_0\rangle |^2
  \zeta(\omega - \omega_{\nu})
\label{eq:strength_function}
\end{equation}
where
$  \zeta(\omega - \omega_{\nu}) =
  \Delta /[2\pi[(\omega- \omega_{\nu})^2+\frac{\Delta^2}{4}]]$
is a Lorentz weight with the averaging parameter $\Delta$=1 MeV.  Such
averaging width is found optimal for the comparison with experiment
and simulation of broadening effects beyond SRPA (escape widths,
coupling with complex configurations).  Further, $\Psi_0$ is the
ground state, $\nu$ runs over the RPA $K^{\pi}=1^+$ states with
energies $\omega_{\nu}$ and wave functions $\Psi_\nu$.
The operator of spin-flip M1 transition reads in standard notation as
$\hat{M}=  \mu_B \sqrt{\frac{3}{8\pi}}\sum_q
[g^{q}_s {\hat s}_{+}^q + g^{q}_l {\hat l}_{+}^q]$
with spin g-factors $g^{p}_s = 5.58 \varsigma_p$ and
$g^{n}_s = - 3.82 \varsigma_n$
quenched by $\varsigma_p$=0.68 and $\varsigma_n$=0.64.
As we are interested in the spin-flip M1, the orbital response is omitted, i.e.
we put $g^{q}_l=0$. Note that in the experimental data
\cite{exp_48Ca,exp_208Pb,exp_Gd_U,exp_Gd,exp_Shizuma_08} used
for the comparison, the orbital contribution is strongly suppressed.
The strength function
(\ref{eq:strength_function}) is computed directly,  i.e. without calculation
of RPA states $\nu$, which additionally reduces the computation effort
\cite{nest_PRC_02,nest_05_lanl,nest_PRC_06,Petr_PhD}.

More details of SRPA formalism are given in the appendices A, B, and C.

\section{Results and discussion}

In Fig. 1 the collective shifts of the main resonance peak in $^{48}$Ca,
$^{208}$Pb, $^{158}$Gd, and $^{238}$U, obtained with different Skyrme
parameterizations, are shown. The shifts are defined as
$E_{shift}=E_{SRPA}-E_{2qp}$, i.e. as a difference in the energies of SRPA and
unperturbed two-quasiparticle M1 peaks. The 2qp strength is calculated by using
(\ref{eq:strength_function}) without the residual interaction.  In addition to
the total shift, the contributions from different spin-density dependent terms
as well as from the tensor force (for SkT6, SG2 and SkO') are shown. It is seen
that the total collective shifts are generally modest and vary from 1-2 MeV in
$^{48}$Ca to 0.5-2 MeV in $^{208}$Pb and 0.5-1.5 MeV in $^{158}$Gd and
$^{238}$U. The low value emerges from contributions pulling in different
directions.  This holds for the separate shifts from $\tilde{b}_0$ and
$\tilde{b}_3$ (not disentangled here). The $\tilde{b}_2$-term gives a negative
shift in contrast to the positive one from $\tilde{b}_0, \tilde{b}_3$. Anyway
the contribution from $\tilde{b}_0, \tilde{b}_3$ usually dominates thus giving
the total upshift in accordance with isovector character of the resonance. All
the forces give generally similar results.  Note a sizable contribution of the
tensor interaction for SkT6 and SG2.  For SkO' this contribution is negligible,
except for $^{48}$Ca where it is so strong that gives a negative total
$E_{shift}$.  It should be emphasized that the non-spin contributions (with
$b_i,b'_i$) alone do not provide any collective shift and leave the M1 strength
unperturbed.  The whole shift is produced by the spin-dependent terms
$\propto\tilde{b}_i,\tilde{b}'_i$.

The calculations give a reasonable summed B(M1) strength. In the interval
0-45 MeV the unperturbed strength is 3.2 and 18.4-18.6 $\mu_N^2$ in $^{48}$Ca and
$^{208}$Pb. The residual interaction changes these values and we have
2.5 - 4.8 $\mu_N^2$ in $^{48}$Ca and 14.8 - 17.3 $\mu_N^2$ in
$^{208}$Pb as compared with experimental values $\sim$ 5.3 $\mu_N^2$
\cite{exp_48Ca} and $\sim$ 17.9 $\mu_N^2$ \cite{exp_208Pb}, respectively. Note
a strong collective effect in $^{48}$Ca.

\begin{figure}[t]
\includegraphics[height=8.5cm,width=6.5cm,angle=-90]{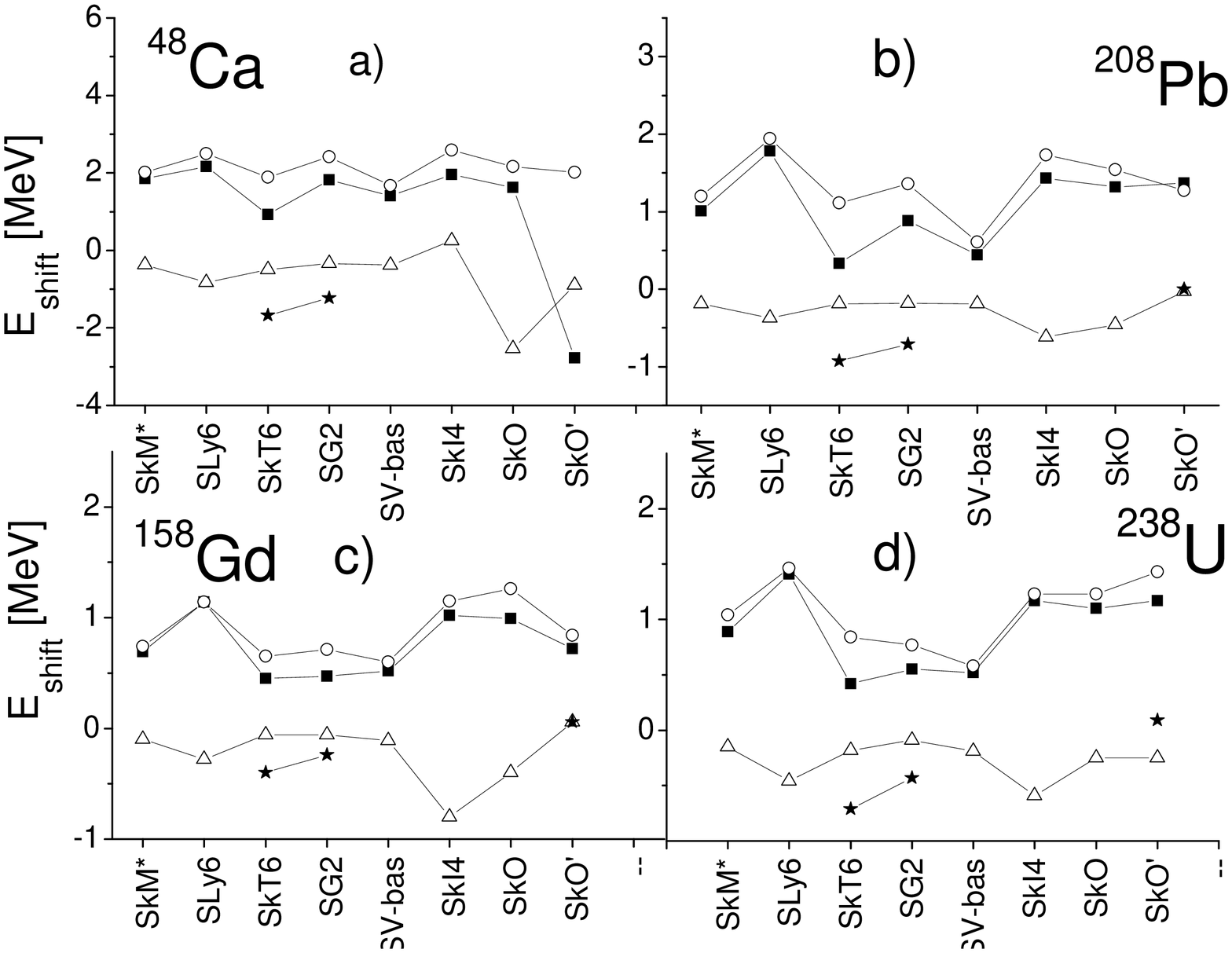}
\caption{ \label{fig:fig1_coll_shifts}
Collective shifts of the M1 peak for different Skyrme
parameterizations (as indicated along the $x$-axis) in $^{48}$Ca,
$^{208}$Pb, $^{158}$Gd and $^{238}$U. The plots exhibit the total
(black boxes) shifts as well as the partial ones with
$\tilde{b}_0$ and $\tilde{b}_3$ (open circles),
$\tilde{b}_2$ (open triangles), and $\tilde{b}_1$
(stars). The $\tilde{b}_1$ contribution exists only for SkT6, SG2, and SkO'.
For better view the symbols are connected by lines.
}
\end{figure}
\begin{figure}[t]
\vspace{0.4cm}
\includegraphics[height=8cm,width=6cm,angle=-90]{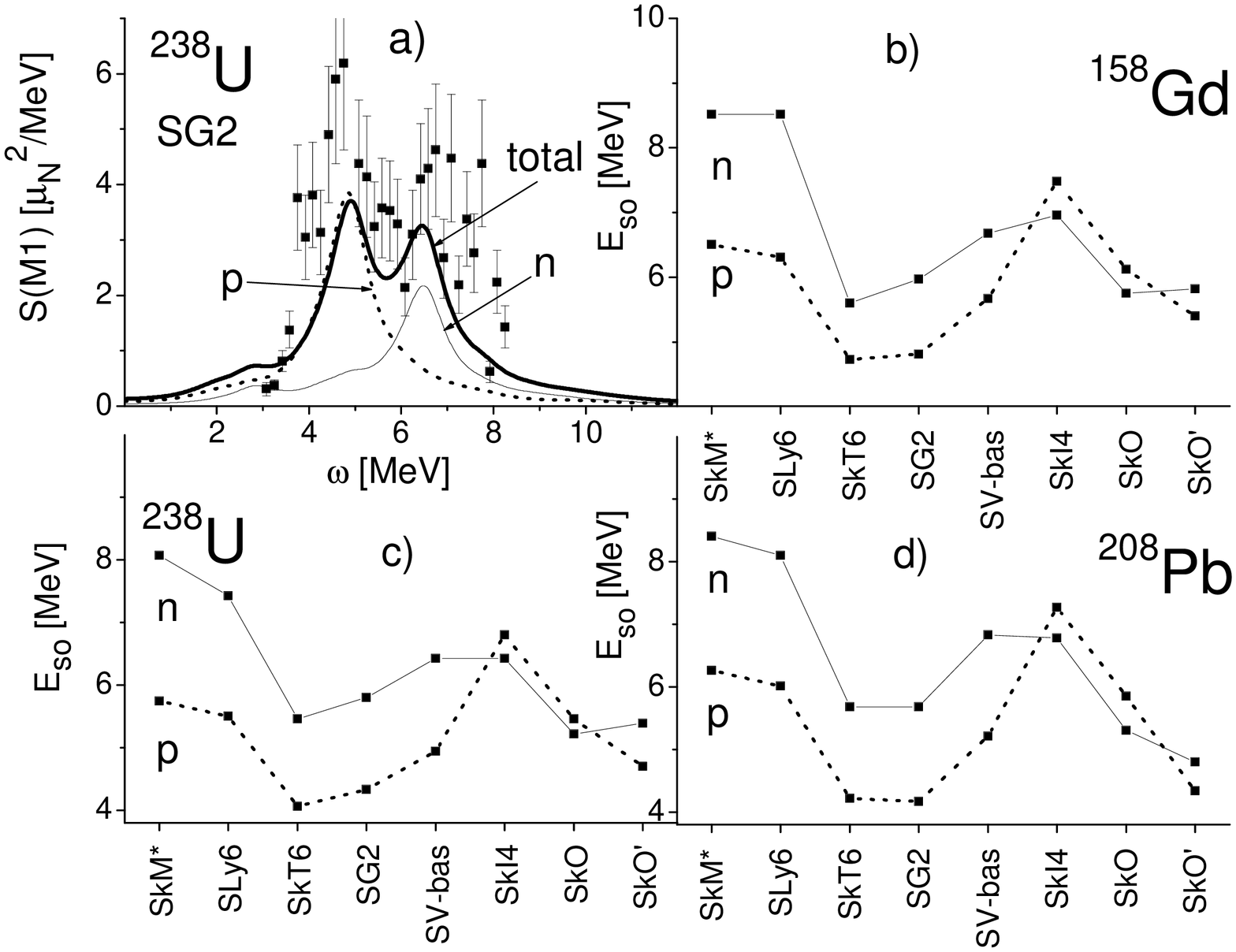}
\caption{ \label{fig:fig2_np}
a): The spin-flip M1 resonance in $^{238}$U calculated with SG2 force:
total (bold solid line), proton (dotted line) and neutron (solid line)
strengths. The experimental data \protect\cite{exp_Gd_U} are
given by black boxes with bars.
b)-d): proton and neutron  unperturbed spin-orbit splittings for different
Skyrme force in  $^{158}$Gd, $^{238}$U, and $^{208}$Pb.
}
\end{figure}
\begin{figure}[t]
\includegraphics[height=8.5cm,width=6.5cm,angle=-90]{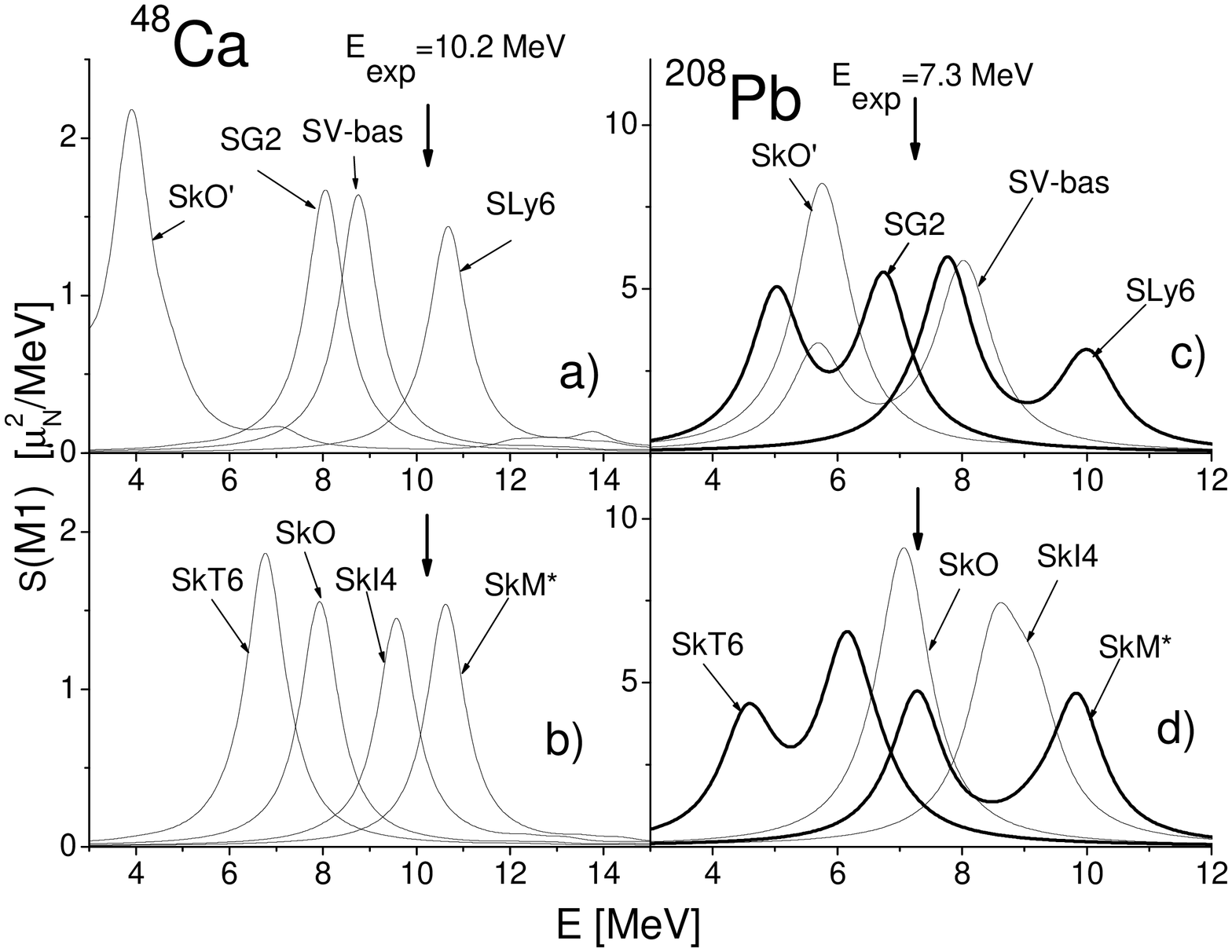}
\caption{ \label{fig:fig3_Ca_Pb_M1}
The spin-flip M1 resonance in $^{48}$Ca (left) and $^{208}$Pb (right)
calculated within SRPA for 8 Skyrme forces as indicated.
For better distinguishability,  the strength in $^{208}$Pb for
SG2, SLy6, SkT6 and SkM* is depicted by the bold line.
The experimental energies in $^{48}$Ca
\protect\cite{exp_48Ca} and $^{208}$Pb \protect\cite{exp_208Pb} are
marked by vertical arrows.
}
\end{figure}

It is well known that proton and neutron spin-orbit splittings
$E^q_{so}$ represent another crucial ingredient in description of
spin-flip M1 mode \cite{Speth_91,Osterfeld_92,Harakeh_book_01}.
Usually $E^p_{so}<E^n_{so}$ which leads in rare-earth and actinide
nuclei to a two-peak structure of the resonance with
dominant proton (neutron) origin of the lower (upper) peak. This is
demonstrated for $^{238}$U in Fig. 2a) where the proton and neutron
components of the M1 mode, obtained with $\varsigma_p$=0.68,
$\varsigma_n$=0 and $\varsigma_p$=0, $\varsigma_n$=0.64, respectively,
are shown.  Panels b)-d) exhibit proton and neutron
splittings $E^q_{so}$ for different Skyrme forces. The
splittings are evaluated from centroids of the proton and neutron
peaks of the unperturbed M1 strength. The results strongly depend on the
parameterization. In most of the cases we have $E^p_{so}<E^n_{so}$
with $E^{np}_{so}=|E^p_{so}-E^n_{so}| \sim $1-2 MeV but SkI4, SkO and
SkO' give very close splittings with even $E^p_{so} > E^n_{so}$ for
SkI4 and SkO. The latter is related with a low value of $b_4$ and
nonzero value of $b'_4$ in SkI4. Note that the experimental
proton-neutron splitting in M1 gross structure is $\sim 2$ MeV in
$^{158}$Gd and $^{238}$U \cite{exp_Gd_U,exp_Gd} and zero in $^{208}$Pb
\cite{exp_208Pb,exp_Shizuma_08}.

In Fig. 3 the SRPA M1 strength function (\ref{eq:strength_function})
in spherical doubly-magic nuclei $^{48}$Ca and $^{208}$Pb is presented.
In $^{48}$Ca, the resonance is produced only by neutron spin-flip transition
$\nu(1f_{7/2}^{-1}, 1f_{5/2})$ yielding
the one-peak structure.  This feature is correctly reproduced by
all the parameterizations. However, most of them underestimate
the resonance energy (with worst SkO' case because of the strong
and possibly wrong tensor contribution) and only the forces  SLy6, SkI4,
and SkM* (with maximal $E^n_{so}$)  give the M1 energy close to
experiment. The success of these forces is obviously determined by a
suitable neutron spin-orbit splitting.
\begin{figure}[t]
\includegraphics[height=8.5cm,width=5.8cm,angle=-90]{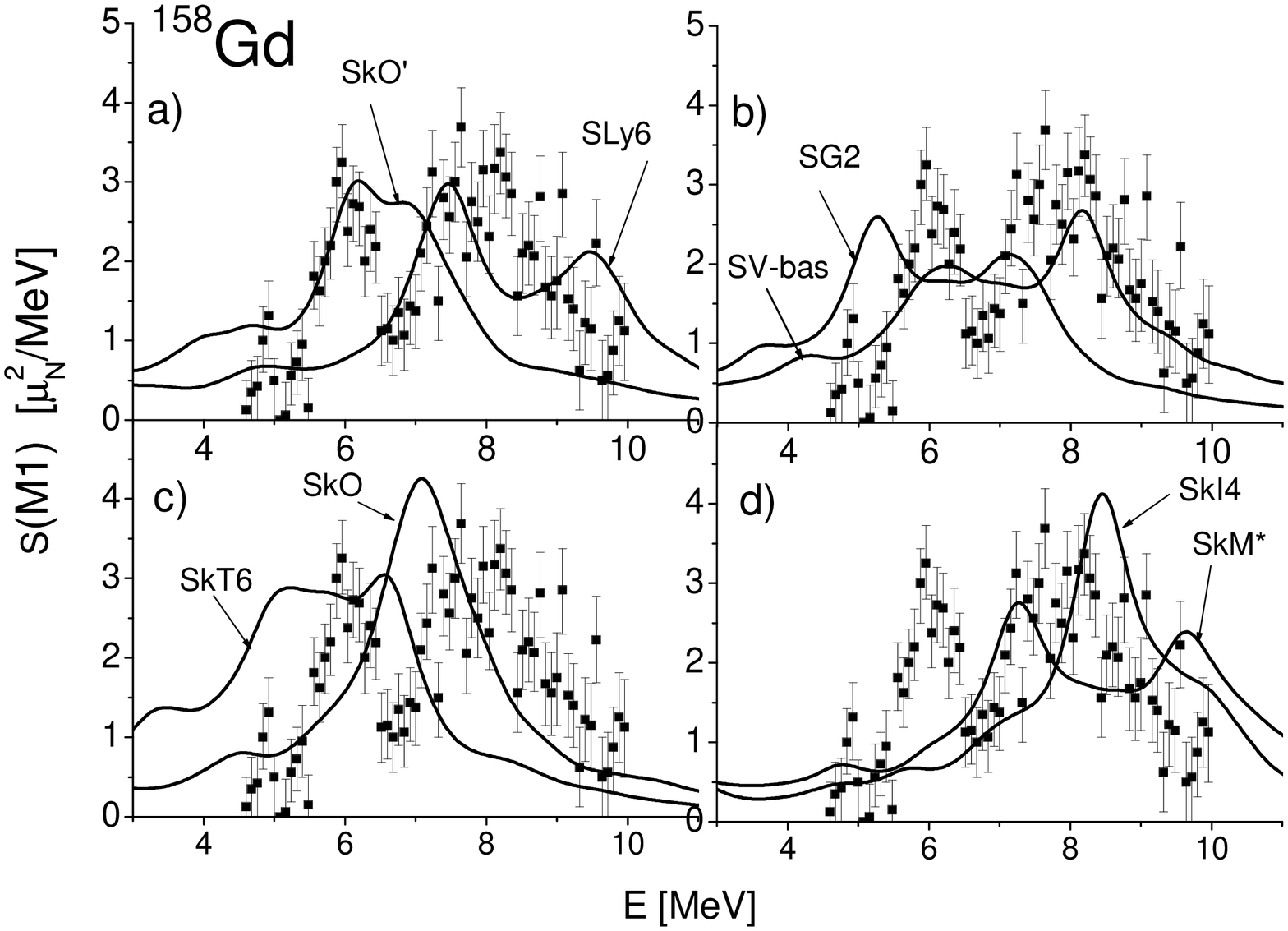}
\vspace{0.6cm}
\caption{ \label{fig:fig4_Gd_U_M1}
The spin-flip M1 resonance in $^{158}$Gd
described with eight Skyrme parameterizations as indicated.
The experimental data are from \protect\cite{exp_Gd,exp_Gd_U}.
}
\end{figure}

However, the same figure shows that in $^{208}$Pb the
forces SLy6, SkI4, and SkM* considerably overestimate the M1
energy while the best result is achieved by SkO. Note that only SkO,
SkO' and SkI4, all having a small $E^{np}_{so}$, give a one-peak
resonance structure in accordance with experiment
\cite{exp_208Pb}. This is because only for these parameterizations
the interaction energy ( $\approx$ collective shift)  is larger
than $E^{np}_{so}$ and so a significant mixture of proton and neutron
components with forming of a one-peak resonance becomes possible.
The other forces have too large $E^{np}_{so}$ and produce the two-peak
structure. This demonstrates the great importance of the interplay
between the residual interaction and relative proton and neutron
spin-orbit splitting $E^{np}_{so}$ for the description of spin-flip
M1.
\begin{figure}[t]
\vspace{0.4cm}
\includegraphics[height=8.5cm,width=6.5cm,angle=-90]
{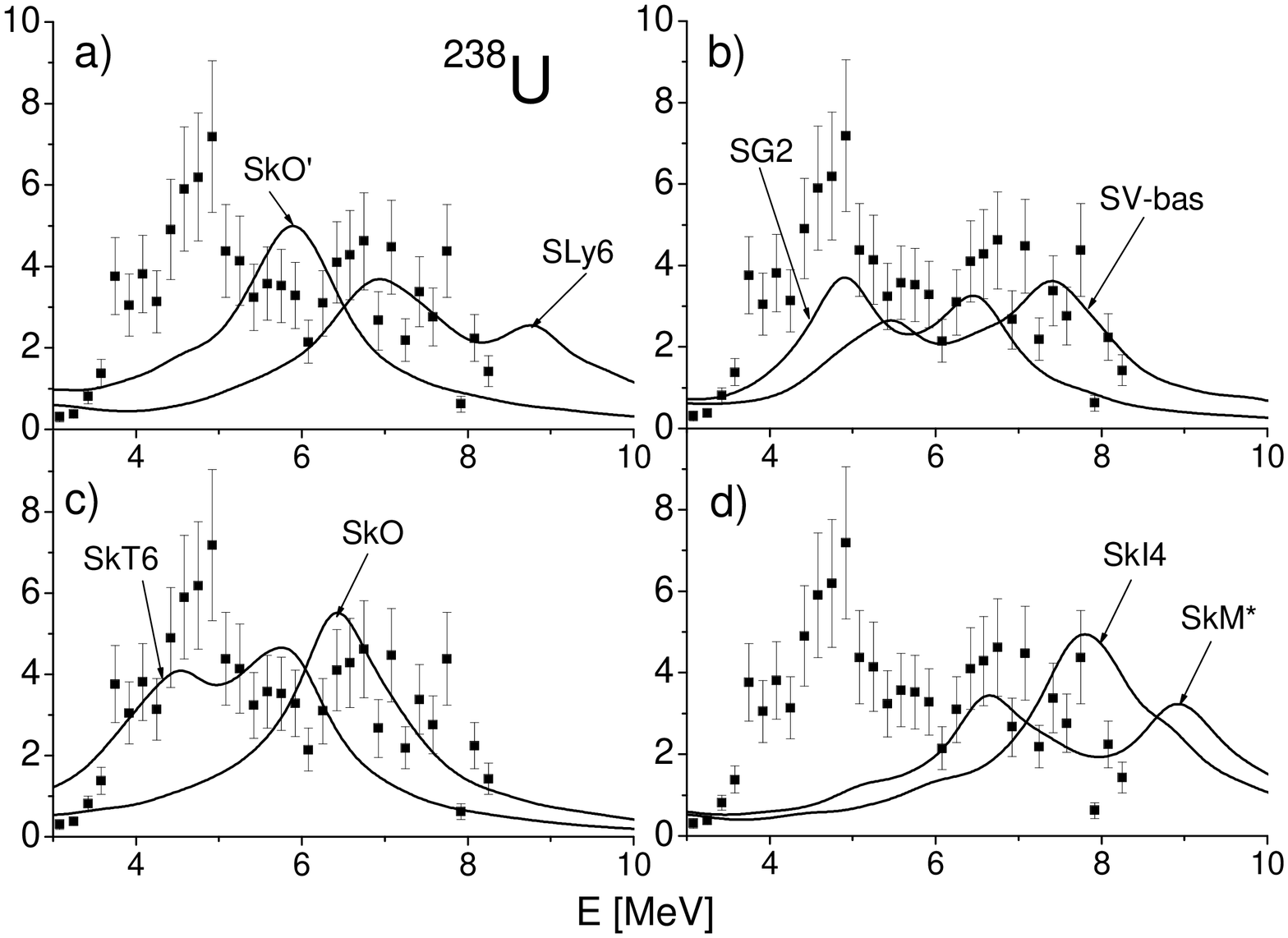}
\vspace{0.8cm}
\caption{
\label{fig:fig4_Gd_U_M1} The same as in Fig. 4 for $^{238}$U.
}
\end{figure}

Figs. 4 and 5 present SRPA results for deformed $^{158}$Gd and
$^{238}$U.  Here, in contrast to $^{208}$Pb, the experiment yields a
double-peak structure and so the one-peak picture from SkO, SkO', and
SkI4 fails. The description is generally quite poor, with exception of
SV-bas in $^{158}$Gd and SG2 in $^{238}$U.
Thus we see that any Skyrme parameterization fails to
describe simultaneously the one-peak structure in closed-shell nuclei
and two-peak structure in open-shell nuclei. The reason
is yet unclear. However, the relative spin-orbit
splitting $E^{np}_{so}$ is evidently one of the key factor and it
is important to find a way to control it.
\begin{figure}
\vspace{0.4cm}
\includegraphics[height=5.5cm,width=7.0cm]{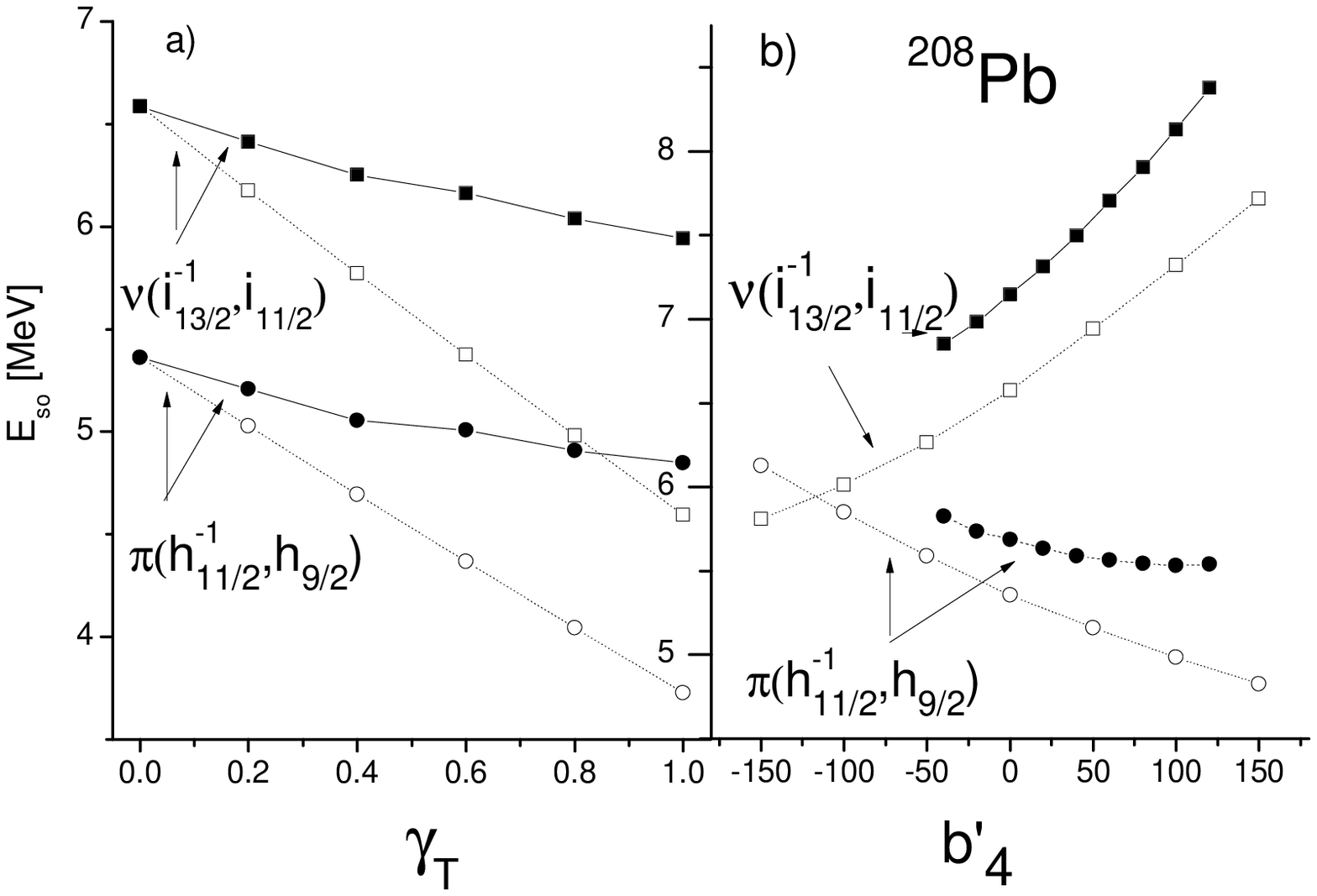}
\vspace{0.3cm}
\caption{ \label{fig:fig5_ten_ils} Dependence of unperturbed
spin-orbit splitting for proton $\pi(h_{11/2}^{-1},h_{9/2})$ and neutron
$\nu(i_{13/2}^{-1},i_{11/2})$ configurations in $^{208}$Pb on (a) the
attenuation $0 \le \gamma_{T} \le 1$ for tensor interaction and (b) the
parameter $b'_4$ of isovector spin-orbit interaction. Skyrme parameterizations
with varied $\gamma_{T}$ or $b'_4$ are used \protect\cite{svbas}. The proton
(neutron) splittings are marked by circles (boxes). Filled symbols mark results
for parameterizations refitted for given $\gamma_{T}$ or $b'_4$. Open symbols
stand for SV-bas where only $\gamma_{T}$ or $b'_4$ are varied, i.e. without
refitting. }
\end{figure}

Fig. 6 explores the influence of the spin-orbit contributions by
systematic variation of tensor spin-orbit (panel a) and isovector
spin-orbit terms (panel b).  Variation of the tensor spin-orbit
strength shifts significantly both $E^{n}_{so}$ and $E^{p}_{so}$ while
leaving the relative order unchanged. But the variation of the
isovector spin-orbit strength $b'_4$ has a strong effect on the
relative positions.  So by simultaneous monitoring tensor and
isovector spin-orbit interactions one may control better the
spin-orbit splittings. Besides the strong effect on single-particle
energies, these interactions also affect the collective shifts
(see Fig. 1). Altogether they represent very promising tool for
further improvement of the description of spin-flip M1 modes.

\section{Conclusions}

We have studied the ability of Skyrme forces to describe the spin-flip M1
resonance using RPA with self-consistent factorized residual interaction. The
results show that the terms with spin and spin-orbit densities are responsible
for a sizable collective shift of the resonance peak. The spin-orbit splitting
of the underlying two-quasiparticle states is of crucial importance for the
final pattern of the spectrum (single-peak versus double-peak structure). The
residual interaction tends to mix of proton and neutron spin-orbit partners and
so works towards a one-peak structure, as in $^{208}$Pb. However, a large
difference between neutron and proton spin-orbit splitting inhibits this mixing
and produces two distinct proton and neutron peaks, as seen experimentally in
rare-earth and actinide nuclei.

None of eight Skyrme parameterizations used in the present study is
able to describe simultaneously the one-peak and two-peak structures in
closed-shell and open-shell nuclei.  In most of the cases, the
resonance energies are badly reproduced as well.  A first exploration
indicates that fine-tuning of
tensor and isovector spin-orbit interactions could improve
the description, as they affect both the spin-orbit splittings and
collective shifts. Work in this direction is in progress. A
corresponding improvement
of the Skyrme parameterizations would be important not only for
description of spin-flip M1 resonance (which is a challenge itself)
but also for better treatment of the spin-orbit interaction
in nuclei.

\section*{Acknowledgments}
The work was partly supported  by the grants DFG-322/11-1, Heisenberg-Landau
(Germany - BLTP JINR), and Votruba - Blokhintsev (Czech Republic - BLTP JINR).
W.K. and P.-G.R. are grateful for the BMBF support under contracts 06 DD 139D
and 06 ER 808. Being a part of the research plan MSM 0021620859 (Ministry of
Education of the Czech Republic) this work was also funded by Czech grant
agency (grant No. 202/06/0363). V.Yu.P. thanks the DFG contract SFB 634.

\appendix 
\section{SRPA}
\label{sec:model}

The SRPA formalism is given elsewhere
\cite{nest_PRC_02,nest_05_lanl,nest_PRC_06}. So here we will
present only principle points and peculiarities pertinent to magnetic
excitations. The SRPA approximates the residual interaction of Skyrme RPA in
a factorized (separable) form as
\begin{equation} \label{V_sep}
  \hat{V}_{\rm res}^{\rm sep}
  =
  -\frac{1}{2}\sum_{qq'}\sum_{k, k'=1}^{K}
   \{ \kappa_{qk,q'k'} {\hat X}_{qk} {\hat X}_{q'k'}
     + \eta_{qk,q'k'} {\hat Y}_{qk} {\hat Y}_{q'k'} \}
\end{equation}
where the indices $q$ and $q'$ label neutrons and protons, $k$ numbers the
separable terms, ${\hat X}_{qk}$ and ${\hat Y}_{qk}$ are time-even and
time-odd hermitian one-body operators, $\kappa_{qk,q'k'}$ and $\eta_{qk,q'k'}$ are
the corresponding strength matrices.  We need these two kinds of the operators
since the relevant Skyrme functionals involve both time-even and time-odd
 densities, see \cite{Ben,nest_PRC_02,nest_05_lanl,nest_PRC_06,Petr_PhD}.

The starting point is a Skyrme functional
$E[J_{q}^{\alpha}({\textbf{r}},t)]=\int d\textbf{r} \mathcal{H}_\mathrm{Sk}(\textbf{r},t)$,
say that in (\ref{eq:skyrme_funct}), with a set of local densities
$J_{q}^{\alpha}$.
The separable operators and strength matrices are derived from the functional
by using the  scaling transformation for the perturbed wave function of the
system:
\begin{equation}\label{eq:scaling}
  |\Psi(t) \rangle_q = \prod_{k=1}^K
  \exp [-i\varrho_{q k}(t)\hat{P}_{q k}]
  \exp [-ip_{q k}(t)\hat{Q}_{q k}]
  | \rangle_q
\end{equation}
were both  $|\Psi(t)\! \rangle_q $ and ground state $| \rangle_q$ are Slater
determinants, ${\hat Q}_{qk}(\textbf{r})$ and ${\hat P}_{qk}(\textbf{r})$ are
generalized coordinate (time-even) and momentum (time-odd) hermitian one-body
input operators, $\hat{H}$ stands for the full Hamiltonian. For E$\lambda$
modes, the input operators ${\hat Q}_{qk}(\textbf{r})$ determine ${\hat
P}_{qk}(\textbf{r})=i[\hat{H},\hat{Q}_{qk}]$, while for M$\lambda$ modes, the
input operators ${\hat P}_{qk}(\textbf{r})$ gives ${\hat
Q}_{qk}(\textbf{r})=i[\hat{H},\hat{P}_{qk}]$. Further
\begin{equation}
\label{eq:q_p} \varrho_{qk}(t)=\bar{\varrho}_{q k} \cos(\omega t) \; ,
p_{qk}(t)=\bar{p}_{qk} \sin(\omega t)
\end{equation}
are corresponding collective variables. As seen from (\ref{eq:scaling}), the
input operators are determined up to arbitrary constant multipliers whose
action can be compensated by the proper rescaling the collective variables.

The number $K$ of input operators in (\ref{eq:scaling}) determines the number
of the separable terms in $\hat{V}_{\rm res}^{\rm sep}$. The treatment
converges to exact RPA  for $K\to\infty$. In practice, a good approximation to
RPA is already obtained for a small $K = 2\div 5$ if the input operators
$\hat{Q}_{q k}$ ($\hat{P}_{q k}$) are properly chosen.

The separable operators and (inverse) strength matrices in (\ref{V_sep}) are
self-consistently constructed as \cite{nest_PRC_02,nest_05_lanl,nest_PRC_06}
\begin{eqnarray}
\label{eq:X}
\hat{X}_{q k} &=& \sum_{q'}\hat{X}_{q k}^{q'} = i\sum_{\alpha' \alpha q'}
\frac{\delta^2 E} {\delta J_{q'}^{\alpha'} \delta J_{q}^{\alpha}} \langle
[\hat{P}_{q k} ,{\hat J}_{q}^{\alpha}] \rangle {\hat J}_{q'}^{\alpha'},
\\
\label{eq:Y} \hat{Y}_{q k} &=& \sum_{q'}\hat{Y}_{q k}^{q'} = i\sum_{\alpha'
\alpha q'} \frac{\delta^2 E} {\delta J_{q'}^{\alpha'} \delta J_{q}^{\alpha}}
\langle [\hat{Q}_{q k} ,{\hat J}_{q}^{\alpha}] \rangle {\hat J}_{q'}^{\alpha'},
\end{eqnarray}
\label{eq:kappa-eta}
\begin{eqnarray}
\label{eq:kappa}
  \kappa_{q'k',qk}^{-1 }
  &=&
  \sum_{\alpha \alpha'}
  \frac{\delta^2 E}{\delta J_{q'}^{\alpha'}\delta J_{q}^{\alpha}}
  \langle [\hat{P}_{q k},{\hat J}_{q}^{\alpha}] \rangle
  \langle [\hat{P}_{q' k'},{\hat J}_{q'}^{\alpha'}] \rangle ,
\\
\label{eq:eta}
  \eta_{q'k',qk}^{-1 }
  &=&
 \sum_{\alpha \alpha'}
  \frac{\delta^2 E} {\delta J_{q'}^{\alpha'}\delta J_{q}^{\alpha}}
  \langle [\hat{Q}_{q k},{\hat J}_{q}^{\alpha}] \rangle
  \langle [\hat{Q}_{q' k'},{\hat J}_{q'}^{\alpha'}] \rangle
\end{eqnarray}
where $\hat{J}_{q}^{\alpha}$ are the density operators.

The final RPA equations have the form
\begin{eqnarray}
\label{eq:RPA_1} \sum_{qk} \{ \bar{\varrho}_{qk}^{\nu} [F_{q'k',qk}^{(XX)}-
\kappa_{q'k',qk}^{-1}] +\bar{p}_{qk}^{\nu} F_{q'k',qk}^{(XY)} \} &=& 0 ,
\\
\label{eq:RPA_2} \sum_{qk} \{ \bar{\varrho}_{qk}^{\nu} F_{q'k',sk}^{(YX)}
+\bar{p}_{qk}^{\nu} [F_{q'k',qk}^{(YY)} - \eta_{q'k',qk}^{-1}] \} &=& 0
\end{eqnarray}
with
\begin{equation}
\label{eq:F_AB}
  F_{q'k',qk}^{(AB)}
 = 2 \sum_{q", ph \in q"}\alpha_{AB}
\frac{\langle ph|\hat{A}^{q"}_{qk} |\rangle^*
   \langle ph|\hat{B}^{q"}_{q'k'} |\rangle}
   {\varepsilon_{ph}^2-\omega_{\nu}^2}
\end{equation}
and
\begin{equation}
\alpha_{AB}=\left(
\begin{array}{l}
{\varepsilon_{ph}, \quad\mbox{for} \; \hat{A}=\hat{B}}\\
{-i\omega_{\nu}, \; \mbox{for} \; \hat{A}=\hat{Y}, \hat{B}=\hat{X}}\\
{i\omega_{\nu}, \quad \mbox{for} \; \hat{A}=\hat{X}, \hat{B}=\hat{Y}}
\end{array}
\right) \; .
\end{equation}
Here $\langle ph|\hat{A}^{q"}_{q'k'} |\rangle$ is the matrix element for the
two-quasiparticle state $|ph\rangle$, $\varepsilon_{ph}$ is the energy of this
state, $\omega_{\nu}$ is the energy of the RPA state $|\nu\rangle$. The
amplitudes of the RPA phonon operator
\begin{equation}
  \hat{C}^{\dagger}_{\nu}
  =
  \sum_q \sum_{ph\in q}
  \left(c^{\nu -}_{ph}\hat{A}^{\dagger}_{ph}
   - {c^{\nu +}_{ph}}_{\mbox{}}\hat{A}^{\mbox{}}_{ph}\right)
\label{eq:geneigen}
\end{equation}
are determined via solutions of (\ref{eq:RPA_1})-(\ref{eq:RPA_2})
\begin{eqnarray}
  c^{\nu \pm}_{ph \in q}
  =
  -\sum_{q'k'}
  \frac{\bar{\varrho}_{q'k'}^{\nu}
        \langle ph|\hat{X}^q_{q'k'}\rangle
        \mp i
        \bar{p}_{q'k'}^{\nu}  \langle ph|\hat{Y}^q_{q'k'}\rangle}
  {2(\varepsilon_{ph}\pm\omega_{\nu})}
\label{eq:c_pm_qp}
\end{eqnarray}
and $\hat{A}^\dagger_{ph}$ and $\hat{A}_{ph}$ are operators of creation and
destruction of two-quasiparticle states.

Following (\ref{eq:X})-(\ref{eq:eta}), the separable ansatz (\ref{V_sep})
explores the residual interaction of the Skyrme functional through the second
functional derivatives. The calculations show that for spin-flip magnetic modes
the spin
\begin{eqnarray}
\frac{\delta^2 E}{\delta \textbf{s}_{q'}(\textbf{r}') \delta
\textbf{s}_{q}(\textbf{r})} &=& \Biggl[ \tilde{b}_0 - \tilde{b}'_0 \delta_{q
q'} + \tilde{b}_3 \frac{2}{3} \rho^{\alpha}(\textbf{r})
\\
- \frac{2}{3} \tilde{b}'_3 \rho^{\alpha}(\textbf{r}) \delta_{q q'}
&-& (\tilde{b}_2 - \tilde{b}'_2 \delta_{q q'}) \Delta_{\textbf{r}} \Biggr]
\delta(\textbf{r}-\textbf{r}') \; , \nonumber
\end{eqnarray}
spin-orbit
\begin{eqnarray}
\frac{\delta^2 E}{\delta \textbf{J}_{q'}(\textbf{r}') \delta
\rho_{q}(\textbf{r})} &=& (b_4 + b'_4 \delta_{q q'}) \nabla_{\textbf{r}}
\delta(\textbf{r}-\textbf{r}')
\; ,
\\
\frac{\delta^2 E}{\delta \textbf{j}_{k;q'}(\textbf{r}') \delta
\textbf{s}_{l;q}(\textbf{r})} &=& (b_4 + b'_4 \delta_{q q'}) (\varepsilon_{klm}
\nabla_{m;\textbf{r}}) \delta(\textbf{r}-\textbf{r}') \; , \nonumber
\end{eqnarray}
and tensor terms
\begin{eqnarray}
\frac{\delta^2 E}{\delta \textbf{J}_{q'}(\textbf{r}') \delta
\textbf{J}_{q}(\textbf{r})} &=& - 2(\tilde{b}_1 + \tilde{b}'_1 \delta_{q q'})
\delta(\textbf{r}-\textbf{r}') \; ,
\\
\frac{\delta^2 E}{\delta \textbf{T}_{q'}(\textbf{r}') \delta
\textbf{s}_{q}(\textbf{r})} &=& (\tilde{b}_1 + \tilde{b}'_1 \delta_{q q'})
\delta(\textbf{r}-\textbf{r}')
\end{eqnarray}
are most important.

SRPA equations presented above are obtained for arbitrary functionals
$E[J_{q}^{\alpha}({\textbf r},t)]$, including Skyrme ones. The model is
self-consistent in the sense that both the static mean field
\begin{eqnarray}\label{eq:h_0}
  \hat{h}_0 =
  \sum_{\alpha q}
  \frac{\delta E} {\delta J_{q}^{\alpha}}\hat{J}_{q}^{\alpha}
\end{eqnarray}
and the residual interaction (\ref{V_sep}), (\ref{eq:X})-(\ref{eq:eta}) are
derived from the same functional. The rank of the RPA matrix
(\ref{eq:RPA_1})-(\ref{eq:RPA_2}) is determined by the number $K$ of the input
operators in (\ref{eq:scaling}). As was mentioned above, usually $K=2 \div 5$
and so the rank is very
small \cite{nest_PRC_06,nest_IJMPE_07,nest_IJMPE_08,nest_PRC_08}. This
drastically simplifies RPA computational effort and allows to perform
systematic explorations even for heavy deformed nuclei.

\section{SRPA strength function}

Giant resonances in heavy nuclei are formed by many RPA states whose detailed
contributions cannot be resolved experimentally. Then, instead of solution of
Eqs. (\ref{eq:RPA_1})-(\ref{eq:RPA_2}), a direct computation of the strength
function (\ref{eq:strength_function}) is more efficient and reasonable. In SRPA
it reads
\begin{eqnarray}
\label{eq:strength_func}
  S(M1 ; \omega) &=& \sum_{\nu \ne 0}
  |\langle\Psi_\nu|\hat{M}|\Psi_0\rangle|^2
  \zeta(\omega - \omega_{\nu})
\\
\nonumber &=&
  \Im\left[
   \frac{z^{L}\sum_{\beta \beta'}
            F_{\beta \beta'}(z) D_{\beta}(z) D_{\beta'}(z)}
        {\pi F(z)}
  \right]_{z=\omega\!+\!i\Delta /2}
\\
\nonumber &+& \sum_q \sum_{ph \in q} \varepsilon_{ph} \langle ph|\hat{M}
|\rangle^2 \zeta(\omega -\varepsilon_{ph})
\end{eqnarray}
where $\beta=qk\tau$ with $\tau$ being the time parity, $\Im$ means the
imaginary part of the value inside the brackets,  $F(z)$ is the determinant of
the RPA matrix (\ref{eq:RPA_1})-(\ref{eq:RPA_2})
 with $\omega_{\nu}$ replaced by the complex
argument $z$, $F_{\beta \beta'}(z)$ is the algebraic supplement of the
determinant, and
\begin{subequations}
\begin{equation}
\label{eq:A_X_1} D_{qk}^{(X)}(z) = \sum_{q'} \sum_{ph \in {q'}}
\frac{\omega_{\nu}
      \langle ph|X^{q'}_{qk} |\rangle \langle ph|\hat{M} |\rangle}
      {\varepsilon_{ph}^2-z^2} \; ,
\end{equation}
\begin{equation}
\label{eq:A_Y_1} D_{qk}^{(Y)}(z) = i \sum_{q'} \sum_{ph \in q'}
\frac{\varepsilon_{ph}
       \langle ph|Y^{q'}_{qk} |\rangle \langle ph|\hat{M} |\rangle}
      {\varepsilon_{ph}^2-z^2} \; .
\end{equation}
\end{subequations}
In the present study we are interested only in spin-flip M1 mode. So the
strength function is calculated only  for $\mu=1$ branch of magnetic dipole
excitations. The branch with $\mu=0$ is omitted as it does not support pure
spin-flip transitions but only its mixture with orbital modes.

\section{Input operators}

The SRPA formalism itself does not provide the input operators ${\hat
P}_{qk}(\textbf{r})$ in the scaling transformation (\ref{eq:scaling}). At the same
time, their choice is crucial to converge the approximate residual interaction
$\hat{V}_{\rm res}^{\rm sep}$ to the true Skyrme one with a minimal number of
the separable terms. We achieve this aim by using ${\hat
P}_{qk}(\textbf{r})$-inputs which compel the separable operators $\hat{X}_{q
k}({\textbf r})$ and $\hat{Y}_{q k}({\textbf r})$ to have maxima in different spatial
regions of the nucleus, both in the surface and interior. The analysis shows
that this way indeed allows to get good convergence already with a few
separable terms.

The physical arguments suggest that the leading scaling operator ${\hat
P}_{q1}(\textbf{r})$ should have the form of the applied external field in the
long-wave approximation, in our case of magnetic field of multipolarity
$\lambda\mu=11$. In present calculations we take decoupled spin and orbital
input operators $\hat{P}_{q1}=\hat{s}^q_+$ and  $\hat{P}_{q2}=\hat{l}^q_+$ and
supplement them by $\hat{P}_{q3}=r^2\hat{s}^q_+$ and
$\hat{P}_{q4}=r^2\hat{l}^q_+$. Then altogether we have K=4 input operators
and  the corresponding operators $\hat{X}_{q k}({\textbf r})$ and
$\hat{Y}_{q k}({\textbf r})$
have maxima in both surface and interior of the nucleus.

In deformed nuclei we should take into account the coupling between magnetic
and electric $K^{\pi}=1^+$ states. So the quadrupole input operator $Q_{q5}=r^2
Y_{21}(\Theta)$ with the counterpart ${\hat P}_{q5}=i[\hat{H},\hat{Q}_{q5}]$ is
added. Then K=5 and we have the RPA matrix of the rank 4K=20.

In terms of two-quasiparticle matrix elements, the relations between input
operators are reduced to
\begin{eqnarray}
\nonumber {\hat Q}_{qk}(\textbf{r})&=&i[\hat{H},\hat{P}_{qk}] \to
\\
\langle ph|\hat{Q}_{qk}|0\rangle &=& 2\varepsilon_{ph} \langle
ph|\hat{P}_{qk}|0\rangle - \langle ph|\hat{X}_{qk}^q|0\rangle
\end{eqnarray}
for magnetic modes and
\begin{eqnarray}
\nonumber {\hat P}_{qk}(\textbf{r})&=&i[\hat{H},\hat{Q}_{qk}] \to
\\
\langle ph|\hat{P}_{qk}|0\rangle &=& 2\varepsilon_{ph} \langle
ph|\hat{Q}_{qk}|0\rangle - \langle ph|\hat{Y}_{qk}^q|0\rangle
\end{eqnarray}
for electric modes.

For more details see
\cite{nest_PRC_02,nest_05_lanl,nest_PRC_06,Petr_PhD}.

\end{document}